\documentclass{iopart}

\begin{document}

 \jl{6}
 
\title[Classical and quantum thermodynamics of horizons]{Classical and Quantum Thermodynamics of horizons in spherically symmetric spacetimes} 
\author{T. Padmanabhan \footnote[1]{E-mail address:  {\tt nabhan@iucaa.ernet.in}}}
\address{Inter-University Center for Astronomy and Astrophysics, Post Bag 4, Ganeshkhind, Pune - 411 007, India}

\begin{abstract}
A general formalism for understanding the thermodynamics of  horizons in spherically symmetric
spacetimes is developed. The formalism reproduces known results in the case of black hole spacetimes
and can
 handle more general situations like: (i) spacetimes which are not
 asymptotically flat
(like the de Sitter spacetime) and (ii) spacetimes with multiple horizons having different temperatures (like the 
Schwarzschild-de Sitter spacetime) and provide a consistent interpretation for temperature, entropy 
and energy. I show that it is possible to write  Einstein's equations
for a spherically symmetric spacetime in the form $TdS-dE=PdV$ near {\it any} horizon
of radius $a$ with $S=(1/4)(4\pi a^2), |E| = (a/2)$ and the temperature $T$ determined
from the surface gravity at the horizon. The pressure $P$ is provided by the source of the 
Einstein's equations and $dV$ is the change in the volume when the  horizon
is displaced  infinitesimally.  The same results can be obtained
by evaluating the quantum mechanical partition function {\it  without using Einstein's equations
or WKB approximation for the action}.  Both the classical and quantum analysis
  provide a simple and consistent interpretation of entropy and energy for de Sitter spacetime as well as for $(1+2)$
dimensional gravity. For the Rindler spacetime the entropy per unit transverse area turns out to be $(1/4)$ while the energy is zero. The approach also shows that the  de Sitter horizon ---  like the
Schwarzschild horizon --- is 
effectively one dimensional as far as the flow of information is concerned, while the
Schwarzschild-de Sitter, Reissner-Nordstrom horizons are not. The implications for spacetimes
with multiple horizons are discussed.
\end{abstract}

\maketitle

\section{The need for local description of horizon thermodynamics}
    This paper provides a classical and quantum description of a mini superspace of spacetimes
    which are spherically symmetric. Since the subject of thermodynamics of horizons
   has nearly three decades of history, it is probably best if I begin with a detailed description
   of motivation. Readers interested in concrete results can skip ahead to section \ref{sectwo}.
   
   The best studied spacetimes with horizons are black hole spacetimes 
    \cite{BHLS}, \cite{bhentropy}. In the simplest context of a Schwarzschild black hole of mass $M$,
    one can attribute an energy  $E=M$,  temperature    $T=(8\pi M)^{-1}$  and
    entropy $S=(1/4)(A_H/L_P^2)$ where
   $A_H$ is the area of the horizon and $L_P = (G\hbar /c^3)^{1/2}$ is the Planck length. (Hereafter, I will
   use units with $G=\hbar=c=1$.)  These are clearly related by the thermodynamic identity
   $TdS = dE$, usually called the first law of black hole dynamics. This result has been obtained
   in much more general contexts and  has been investigated from many different points of 
   view in the literature. The simplicity of the result depends on the following features: 
   (a)  The Schwarzschild metric  is a vacuum solution
    with no pressure so that there is no $PdV$ term in the first law of thermodynamics.
    (b) The metric has only one parameter $M$ so that changes in all physical parameters can be related
    to $dM$. (c) Most importantly, there is a well defined notion of energy $E$ to the 
    spacetime and the changes in the energy $dE$ can be interpreted in terms of the 
    physical process of the black hole evaporation. (One can also interpret the relations
    by, say, dropping ``test particles" into the black hole but it is really not necessary.) 
The idea can be generalized to other 
    black hole spacetimes in a rather simple manner only because of well defined notions of
    energy, angular momentum etc.
    
    Can one generalize the thermodynamics of horizons to cases other than black holes
    in a straight forward  way ? In spite of years of research in this field, 
 this generalization remains non trivial and challenging when the conditions
    listed above are not satisfied. 
    To see the importance of the above conditions, we only need to contrast the situation in Schwarzschild 
    spacetime with that of de Sitter spacetime and observe that:
    
    (1)  The notion of temperature is  well defined in the case of de Sitter spacetime since the de Sitter horizon also radiates like a black body;  we have $T= H/2\pi$
    where $H^{-1}$ is the radius of the de Sitter horizon. (The same result can be obtained more formally in terms
    of the periodicity in the Euclidean time). But the correspondence probably
    ends there. 
    A study of literature shows that there exist very  few concrete calculations of energy, entropy
   and laws of horizon dynamics in the case of de Sitter spacetimes, in sharp contrast to
   BH space times. 
   
   (2) There have been several attempts in the literature to define the concept of
   energy using local or quasi-local concepts (for a small sub sample of ten references,
   see  \cite{energydef}). The problem is that not all definitions of energy  (or even definitions of horizon; see \cite{abhay})
   agree with each other
   and not all of them can be applied to de Sitter type universes.  
   
   (3) Even when a notion of 
   energy can be defined, it is not clear  how to
    write and interpret an equation analogous to $dS = (dE/T)$ in this spacetime,
    especially since the physical basis for $dE$ would require
    a notion of evaporation of the de Sitter universe.
    
    (4) Further, we know that de Sitter spacetime is a solution to Einstein's equations with a 
    source having non zero pressure. Hence one would very much doubt whether $TdS $ is 
    indeed equal to $dE$. It would be necessary to add a $PdV$ term for consistency.

   An argument is sometimes advanced that de Sitter horizon is conceptually different
   from black hole horizon because it is observer dependent. I believe this argument is 
   incorrect and that all horizons (including even Rindler horizon) should be treated on par
   because of at least three reasons (see \cite{tprealms} \cite{tpmpla} for more details):
   (i) To begin with, if the notion of entropy in black hole
    spacetimes is not accidental,  then
    one would expect {\it any} one-way-membrane which blocks out information to lead to a 
    notion of entropy \cite{tpmpla}. (ii) As regards  observer dependence, even in the case of Schwarzschild spacetimes,
    it is possible to have observers moving in time-like trajectories inside the event horizon who will
    access part of the information which is not available to the outside observer. It seems unlikely
     that these suicidal observers will attribute the same amount of entropy to the Schwarzschild
    black hole as an observer playing it safe by staying far away from the event horizon. (iii)
   If the notion of entropy associated with a one way membrane arises from
   {\it local} degrees of freedom and Planck scale physics,  our inability to define a {\it global} notion of 
   energy should  not have a bearing on the issue of 
   entropy.

   All these suggest that there must be a local approach
   by which one can define the notion of entropy {\it and} energy for spacetimes with 
   horizons. This conclusion is strengthened further by the following argument:
   Consider a class of spherically symmetric spacetimes of the form
    \begin{equation}
     ds^2=f(r)dt^2-f(r)^{-1}dr^2 - r^2(d\theta^2+\sin^2\theta
     d\phi^2)
     \label{ssmetric}
     \end{equation}
   If $f(r)$ has a simple zero at $r=a$ with $f'(a)\equiv B$ remaining finite, then
   this spacetime has a horizon at $r=a$. 
   Spacetimes like Schwarzschild or de Sitter have only one free parameter
    in the metric (like $M$ or $H^{-1}$) and hence the scaling of all other thermodynamical 
    parameters is uniquely fixed by purely dimensional considerations.  But, for a general  metric
    of the form in (\ref{ssmetric}), with an arbitrary $f(r)$, the 
    area of the horizon  (and hence the entropy) is determined by the location of the zero of the function $f(r)$
    while the temperature --- obtained from the periodicity considerations ---
    is determined by the value of $f'(r)$ at the zero. For a general function, of course,
    there will be no relation between the location of the zero and the slope of the 
    function at that point. It will, therefore, be incredible if there exists any a priori relationship 
     between the temperature (determined by $f'$ ) and the entropy
    (determined by the zero of $f$) even in the context of horizons in spherically symmetric spacetimes. 
    If we take the entropy to be $S=\pi a^2$ (where $f(a)=0$ determines the radius of the horizon) and
    the temperature to be $T=|f'(a)|/4\pi $ (determined by the periodicity of Euclidean time),
    the quantity $TdS  = (1/2) |f'(a)| a da$ will depend both on the slope $f'(a)$ as well as
    the radius of the horizon. This implies that any local interpretation of thermodynamics
    will be quite non trivial.

     Finally, the need for local description of thermodynamics of horizons becomes crucial
     in the case of spacetimes with multiple horizons. Let me briefly describe this situation
     relegating the details to Appendix \ref{appa}.
     The strongest 
   and the most robust result we have, regarding spacetimes with a horizon,
   is the notion of temperature associated with them. This, in turn, depends either on
   a  complicated analysis of the mode functions of a wave equation,
   Bogoliubov coefficients etc. or on the study of the periodicity of the Euclidean time coordinate.  
   Neither approach works very well if the spacetime has more than one horizon like, for example,
  in  the Schwarzschild-de Sitter metric which has the form in (\ref{ssmetric}) with
     \begin{equation}
     f(r) = \left(1-{2M\over r} - H^2 r^2\right)
    \end{equation}
    This spacetime has  two horizons at $r_\pm$ with 
    \begin{equation}
    r_+ =\sqrt{4\over 3} H^{-1} \cos {x+4\pi\over 3};\quad
     r_- =\sqrt{4\over 3} H^{-1}  \cos {x \over 3}
     \end{equation}
     where
     $\cos x =-3\sqrt{3} MH^{-1}$.
     (The parameter $x$ is in the range ($\pi, (3/2)\pi]$ and $0\le 27M^2H^{-2} <1$.)
     Close to either horizon the spacetime can be approximated as Rindler.
      Since the surface gravities on the two
     horizons are different, we get two different Rindler  temperatures $T_\pm = |f' (r_\pm)|/ 4\pi$.
     It is also possible to introduce two {\it different} Kruskal like coordinates 
     using the two surface gravities in which
     the metric is well behaved at the horizons (see \ref{appa} for details; for related work,
see \cite{fengli}).
     In the (1+1) case, it is possible to introduce the analogs of Boulware
     and Hartle-Hawking vacuum states with {\it either} of these transformations.
     In the overlapping region of the coordinate patches, however, there is no simple notion
     of temperature.

     This point
     is brought in quite dramatically when we study the periodicity in the
     imaginary time in the overlapping region of the two Kruskal coordinates.
     To maintain invariance under $it\to it+\beta$ (with some finite $\beta$) it is necessary
     that $\beta$ is an integer multiple of both $4\pi/   |f' (r_+)| $
     and $4\pi/   |f' (r_-)| $ 
     so that $\beta = (4 \pi n_\pm/ |f' (r_\pm)| )$ where $n_\pm$ are integers.
     Hence the ratio of surface gravities $ |f' (r_+)|/ |f' (r_-)| = (n_+/n_-)$
     must be a rational number. 
      Though irrationals can be approximated by rationals,
     such a condition definitely excludes a class of values for 
     $M$ if $H$ is specified and vice versa.
     It is not clear why the existence of a cosmological constant 
     should imply something for the masses of black holes (or vice versa).
    Since there is no physical basis for such a condition,
    it seems reasonable to conclude that these difficulties arise
    because of our demanding the existence of a finite periodicity $\beta$ in the
    Euclidean time coordinate. This demand is related to
    an expectation of thermal equilibrium which is violated in spacetimes with 
    multiple horizons having different temperatures. 
    
 If even the simple notion of temperature falls apart in the presence
     of multiple horizons, it is not likely that the notion of energy or entropy can be 
     defined by {\it global} considerations. 
    On the other hand, it will be equally  strange  if  we cannot attribute a temperature to a black
     hole formed in some region of the universe just because the universe
     at the largest scales is described by a de Sitter spacetime, say.
     One is again led to searching for a {\it local} description of the thermodynamics of 
     {\it all types of}  horizons. 
    
 The results in this paper have bearing on all these important issues.
   One of the key results of this paper will be to provide a consistent interpretation
   of the relation $TdS - dE = PdV$ in the case of de Sitter universe. To the extent I 
   know, this has not been done before in any of the published literature, explicitly, 
   giving an expression for the energy in the de Sitter universe. I obtain  the result
   $E=-(1/2) H^{-1}$ by three separate arguments and these arguments
   work consistently for (1+2) dimensional spacetime as well.
   I will also show [in section \ref{sectwo}] that one can use Einstein's constraint equations to
    relate $f'(r)$ with $r$ and thus provide a simple interpretation for the expression $TdS$.
    [The constraint equation arises essentially from the symmetries of the Einstein-Hilbert
    action under coordinate transformations and can be imposed as a constraint on physical
    states even in quantum theory.]
   
   \section{\label{sectwo} Einstein's equations as a thermodynamic identity}

     The key idea developed in this Section is to use the notion of periodicity in Euclidean time 
(obtained by a local Rindler approximation near the horizon) to define the 
     temperature, without any ambiguity in the proportionality
     constant, and then to rewrite Einstein's equations in a form analogous to the 
     $TdS -dE= PdV$ equation. I will show that it is fairly straight forward to 
     achieve this in the case of spacetimes of the form:
     \begin{eqnarray}
     ds^2&=&f(r)dt^2-f(r)^{-1}dr^2 - r^2(d\theta^2+\sin^2\theta
     d\phi^2)\nonumber \\
     & \equiv& f(r)dt^2-f(r)^{-1}dr^2 -dL_\perp^2
     \label{basemetric}
     \end{eqnarray}
      with a general 
      $f(r)$ determined via Einstein's equations.
     
     This metric  will satisfy Einstein's equations  provided the  the source stress 
     tensor has the form 
     \begin{equation}
     T^t_t = T^r_r ={\epsilon(r)\over 8\pi }; \quad T^\theta_\theta = T^\phi_\phi={\mu(r)\over 8\pi }
     \label{arbfr}
     \end{equation}
     The equality $T^t_t = T^r_r $ arises from our assumption $g_{00}=(-1/g_{11})$ and can 
     be relaxed if  $g_{00}\ne (-1/g_{11})$. (It will turn out that our analysis goes through even in this
     more general case and I will comment about it in the next section.) The equality  $T^\theta_\theta = T^\phi_\phi$
     arises from spherical symmetry. The  equation (\ref{arbfr}) also defines the functions $\epsilon (r) $ and 
     $\mu(r)$.
    The Einstein's equations now reduce to: 
    \begin{equation}
    {1\over r^2} (1-f) - {f'\over r} = \epsilon; \quad \nabla^2 f = -2 \mu
    \label{albert}
    \end{equation}
    The remarkable feature about the metric in (\ref{basemetric}) is that the Einstein's 
    equations become linear in $f(r)$ so that solutions for different $\epsilon(r)$
    can be superposed. (I could not find this result in  standard textbooks.) Given any $\epsilon (r)$ the solution becomes
    \begin{equation}
    f(r) = 1-{a\over r} -{1\over r}\int_{a}^r \epsilon(r)  r^2\, dr
    \label{epsol}
    \end{equation}
    with $a$ being an integration constant and $\mu(r)$ is fixed by $\epsilon(r)$ through:
    \begin{equation}
    \mu(r) = \epsilon + {1\over 2}r\epsilon' (r)
    \label{muofr}
    \end{equation}
    I have chosen the integration constant $a$ in (\ref{epsol})  such that $f(r)=0$ at $r=a$ so that this surface
    is a horizon. It is, of course,
    quite possible for $f(r) $ to vanish at other values of $r$ if there are multiple horizons
    in the spacetime. Given any $\epsilon (r)$, the  two equations (\ref{epsol}) and (\ref{muofr}) give a classical
    solution to Einstein's equations with (at least) one horizon at $r=a$. 
    
    One can easily
    verify that: (i) $\epsilon =0$ implies $\mu=0$ and leads to Schwarzschild spacetime;
    (ii) $\epsilon=\epsilon_0 =$ constant requires $\mu =\epsilon_0$ and 
    $T^a_b \propto {\rm dia} \ (\epsilon_0, \mu_0, \mu_0, \mu_0)$ and leads to
    \begin{equation}
    f(r) = 1 - {A\over r} - Br^2; \quad A=a - {\epsilon_0 a^3\over 3} ;
    \quad B={\epsilon_0\over 3}
    \end{equation}
    This represents the Schwarzschild-de Sitter spacetime and when
    $a=0$ it reduces to the de Sitter spacetime; (iii) $\epsilon=(Q^2/r^4)+\epsilon_0$
    gives $\mu =- (Q^2/r^4)+\epsilon_0$ which corresponds to a Reissner-Nordstrom-de Sitter
    metric. In fact, the linearity of Einstein's equations (\ref{albert}) allows superposition of different
    $\epsilon$ and one can build all these solutions by superposition. 
    
    Let us now assume that the solution  (\ref{epsol}) is such that $f(r) =0$ at $r=a_i, i=1,2,....$ with
    $f'(a_i)=B_i$ finite. Then near $r=a_i$ the metric can be expanded in a Taylor
    series with $f(r) \approx B_i(r-a_i)$. If there are multiple horizons, such
    expansion is possible near each zero of $f$ and I will suppress the subscript $i$ and just denote by
    the symbols $a$ and $B$ the corresponding values for each horizon.
    If only one horizon is present, then the Euclidean time coordinate is periodic
    with a period $\beta =(4\pi/|B|)$ leading to a {\it global} notion of temperature. 
    In the context of multi horizon spacetimes, it is natural to endow each horizon with a
     locally defined temperature  $T_i=|B_i|/4\pi$.
   This can be defined precisely  in the context of multi horizon spacetimes
    by considering the metric near each horizon and identifying the local Rindler 
    temperature or by developing the quantum field theory. (See Appendix \ref{appa};  
    in the case of a multi horizon spacetime, there is, of course,
    no notion of a {\it global} equilibrium temperature.)
    From  the first of the equations (\ref{albert}) evaluated at $r=a$, we get     
    \begin{equation}
    f'(a)={1\over a} - \epsilon(a) a=B
    \end{equation}
    or,      
    \begin{equation}
    {1\over 2} Ba - {1\over 2} =- {1\over 2}\epsilon(a) a^2
    \label{albertmod}
    \end{equation}
    It is possible to provide an interesting interpretation of this equation which 
    throws light on the notion of entropy and energy.
   Multiplying the above equation by
    $da$ and using $\epsilon = 8\pi T^r_r$, it is trivial to rewrite  equation (\ref{albertmod})
    in the form    
    \begin{eqnarray}
    {B\over 4\pi} d\left( {1\over 4} 4\pi a^2 \right) - {1\over 2} da&=&- T^r_r(a) d \left( {4\pi \over 3}  a^3 \right)
    \nonumber \\
    &=&
    - T^r_r(a) [4\pi a^2]da 
    \label{keyeqn}
    \end{eqnarray}
    (It should be remembered that this equation holds separately for each of the horizons with $a,B$ etc. 
    actually standing for $a_i,B_i, i=1,2,...$).
   Let us first consider the case in which a particular horizon has
    $f'(a)= B >0$ so that the temperature is $T=B/4\pi$. Since $f(a)=0,f'(a)>0$, it follows that $f>0$ for
    $r>a$ and $f<0$ for $r<a$; that is, the ``normal region'' in which $t$ is time like is outside
    the horizon as in the case of, for example, the Schwarzschild metric.  
    The first term in the left hand side of (\ref{keyeqn}) clearly has the form of $TdS$ since we have an independent identification of temperature from the periodicity argument in the local Rindler coordinates. 
    Since the pressure is $P=-T^r_r$, 
     the right hand side has the structure of $PdV$ or --- more
    relevantly --- is the product of the radial pressure times the transverse area times the 
    radial displacement. This is important because, for the metrics in the form (\ref{basemetric}), the proper transverse area is just that of a 2-sphere though
    the proper volumes and coordinate volumes differ. The product of pressure $P$ times the transverse {\it proper} area 
$4\pi a^2$ gives the correct force; multiplying this force by the [virtual] displacement $da$ gives the [virtual] work done. Hence  the relevant quantity is $dV=(4\pi a^2) da$ and its integral $V=(4\pi/3)a^3$, sometimes called `areal volume', is the  relevant volume for our analysis. The second equality in (\ref{keyeqn}) makes this point clear. 
[This interpretation is discussed further in subsection (\ref{int}) below.]
    In the case of horizons with $B=f'(a)>0$ which we are considering (with $da>0$), the volume of the region where $f<0$ will increase and the volume of the region
    where $f>0$ will decrease. 
    Since the entropy is due to the existence of an inaccessible region, $dV$ must refer to the change 
    in the volume of the inaccessible region where $f<0$.
    [We will see below that consistent interpretation is possible for $B<0$ case as well.]
    We can now identify
    $T$ in $TdS$ and $P$ in $PdV$ without any difficulty and  interpret the remaining term (second term in the left hand side) as $dE=da/2$.
    We thus get the expressions for the entropy $S$ and energy $E$ (when $B>0$) to be
   \begin{equation}
    S={1\over 4} (4\pi a^2) = {1\over 4} A_H; \quad E={1\over 2} a =\left( {A_H\over 16 \pi}\right)^{1/2}
    \end{equation}
    In the case of the Schwarzschild black hole with $a=2M$, the energy turns out to
    be $E=(a/2) = M$ which is as expected. More generally,
	$E=(A_{\rm horizon}/16\pi)^{1/2}$ corresponds to the so called `irreducible mass' in 
       black hole spacetimes \cite{irrmass}. Of course, the identifications $S=(4\pi M^2)$,
    $E=M$, $T=(1/8\pi M)$ are consistent with the result $dE = TdS$ in this particular case
    since the pressure vanishes. As I said before, reproducing this result for black holes is not
so significant, because much more formal and general results exist in the case of black hole
thermodynamics. What is significant is the fact that our analysis is completely local and did
not use any feature regarding asymptotic flatness etc even to define the energy.

\subsection{Energy of the de Sitter horizon}
    
   This aspect becomes clearer when we study a horizon which is {\it not} associated with a black hole,
viz. de Sitter horizon. There is considerable interest in this spacetime recently but no
clear formulation of ``laws" analogous to laws of black hole dynamics exist in this context (Except possibly
for one approach,
 based on the concept of isolated horizons, that attempts to provide such an analysis \cite{abhay}
using a very specific definition for spacetime horizons.)
Equation (\ref{keyeqn}), however, can easily  provide an interpretation of entropy and energy in the case
    of de Sitter universe. In this case, $f(r) = (1-H^2r^2)$, $a=H^{-1}, B=-2H<0$
    so that the temperature --- which should be positive --- is $T=|f'(a)| / (4\pi) =(-B)/4\pi$.
   For horizons with $B=f'(a)<0$ (like the de Sitter horizon) which we are now considering, 
   $f(a)=0, f'(a)<0$, and it follows that $f>0$ for
    $r<a$ and $f<0$ for $r>a$; that is, the ``normal region'' in which $t$ is time like is inside
    the horizon as in the case of, for example, the de Sitter metric.  
    Multiplying equation (\ref{keyeqn}) by $(-1)$,  we get
     \begin{equation}
    {-B\over 4\pi} d\left( {1\over 4} 4\pi a^2 \right) + {1\over 2} da= T^r_r(a) d \left( {4\pi \over 3}  a^3 \right)
    =P(-dV)
    \label{tdseqn}
    \end{equation}
    The first term on the left hand side is again of the form $TdS$ (with positive temperature and entropy).
    The term on the right hand side has the correct sign since the
     inaccessible region (where $f<0$) is now outside the horizon and the volume of this region 
    changes by $(-dV)$.
   Once again, we can use (\ref{tdseqn}) to identify the entropy and the energy: 
 \begin{equation}
 S={1\over  4} (4\pi a^2) = {1\over  4} A_{\rm horizon}; \quad E=-{1\over 2}H^{-1}
 \end{equation}

As a byproduct, our approach provides an interpretation of energy for the de Sitter spacetime --- an
issue which is currently attracting attention --- and a consistent thermodynamic interpretation of 
de Sitter horizon. Our identification, $E=-(1/2) H^{-1}$ is also supported by the following argument:
If we use the ``reasonable" assumptions $S=(1/4)(4\pi H^{-2}), V\propto H^{-3}$ and $E=-PV$ in the equation
$TdS -PdV =dE$ and treat $E$ as an unknown function of $H$, we get the equation
\begin{equation}
 H^2{dE\over dH}=-(3EH+1)
 \label{arguetwo}
 \end{equation}
which integrates to give precisely $E=-(1/2)H^{-1}. $ 
Note that I only needed the proportionality, $V \propto H^{-3}$ in this argument since $PdV \propto 
(dV/V)$. The ambiguity between the coordinate and proper volume is again irrelevant.
This energy is also
 numerically same as the
total energy within the coordinate Hubble volume of the classical solution, with a cosmological constant:
    \begin{equation}
    E_{\rm Hub}={4\pi \over 3} H^{-3} \rho_\Lambda = {4\pi \over 3} H^{-3} {3H^2\over 8\pi } ={1\over 2}H^{-1}
    \label{normale}
    \end{equation}
  (The extra negative sign of $E=-E_{\rm Hub}$ is related to a feature noticed in the literature
in a different context; see for example, the discussion following equation (71) in the review \cite{negreview}. )

   There is an interesting feature in this argument which is worth discussing (\cite{israel}). 
   If the formal results of black hole dynamics valid for arbitrary matter filled stationary 
    spaces containing a horizon \cite{bch} are carried over to the de Sitter spacetime,
    then one either obtains $E=+(1/2) H^{-1}$ or $E=- H^{-1}$. The first one
    arises essentially due to the result in (\ref{normale}) while the second 
    one arises if the effective mass density is taken to be $(\rho + 3P)$.
    The result obtained above $E = -(1/2)H^{-1}$ differs from both these expressions.
    It is easy to see that the approach based on Einstein's equations as well
    as the argument given above uses $\rho $ as the relevant energy
    density and not $\rho+3P$ and hence differs from the result $E=- H^{-1}$.
    Settling whether $E=+(1/2) H^{-1}$ or $E=-(1/2) H^{-1}$ is more 
    subtle because of the following reason.
   In the case
   of a black hole, increase in the area of horizon $A_H$  leads to increase in the volume of 
   inaccessible region. It therefore makes sense to assume that $(dS/dA_H) >0$.
   In the case of a de Sitter universe, increase in the area of a horizon increases
   the region accessible to the canonical observer in the region $r<H^{-1}$ and thus
   decreases the inaccessible region beyond the horizon. (This notion is not
   very precise since the volume in the region $r>H^{-1}$ could be divergent.)
   From this point of view, it may be acceptable to take $(dS/dA_H) <0$. 
   This will change the sign in $TdS$ term  in the relevant equations and repeating
   our analysis, straightforward algebra will lead to $E=(1/2) H^{-1}$. 
   This is most easily seen from the argument given in the last paragraph. If
   we take $dS = - (1/4) dA_H$, equation (\ref{arguetwo}) changes to
   \begin{equation}
 H^2{dE\over dH}=-(3EH-1)
 \end{equation}
    which has the solution $E=+(1/2) H^{-1}$.
    So, the choice of $E=\pm (1/2) H^{-1}$ is related to the choice of 
    $dS = \mp (1/4) dA_H$. 

I believe the correct result is indeed $E=-(1/2) H^{-1}$
    because we expect total entropy of a system to be a well defined
    positive quantity. 
    Integrating the equation $dS = - (1/4) dA_H$ leads to the result
    $S=(1/4) (A_0 - A_H)$ with an undetermined constant $A_0$.
    This result is hard to interpret since there are no physical quantities
    available to characterize $A_0$ in these spacetimes and the choice
    of $A_0 =0$ will now lead to negative entropy.  There is nothing wrong
    in $dS$ being negative but in a complete thermodynamic description of 
    the spacetime horizon, $S$ has to be positive and should not 
    depend on any arbitrary constant.

While (\ref{keyeqn}) gives a consistent interpretation, one may wonder about the uniqueness. For example,
one could have multiplied the entire equation by an arbitrary function  $F(a)$ which could even differ from
horizon to horizon in the case of multi-horizon spacetimes. In that case (taking $B>0$), the expressions for entropy, energy
and volume will become:
\begin{equation}
S=\int 2\pi F(a) a da; E={1\over 2}\int F(a) da; V=\int 4\pi a^2 F(a)da
\end{equation}
It seems reasonable to assume that the cross section area on which the radial pressure should act must
be $4\pi a^2$ since it is the {\it proper} area in the class of metrics which we are considering. Given this criterion,
it follows that $F(a)=1$. There is no freedom --- not even that of rescaling --- left after this.

\subsection{Results for (1+2) dimensional gravity}

   The ideas also work in the case of $(1+2)$ dimensional gravity which has attracted fair
   amount of attention (see ref. \cite{carlip}). For the metrics in (\ref{basemetric}) with $dL_\perp^2
   = r^2 d\theta^2$, Einstein's equations demand that the stress tensor has the form
   $ 8\pi T^a_b = {\rm dia}\ (\epsilon(r), \epsilon(r), \mu(r))$. The Einstein's equations
   are 
   \begin{equation}
   -{1\over 2} {f'\over r} = \epsilon(r); \quad -{1\over 2} f''(r) = \mu(r)
   \end{equation}
   These are also linear in the source term and can be integrated for a 
   given function $\epsilon(r)$. The solution is 
   \begin{equation}
   f(r) = - 2 \int_a^r dx\ x \epsilon(x); \quad \mu(r) = (r\epsilon(r))'
   \end{equation}
   where we have again chosen the boundary condition such that 
   $f(a)=0$.
   The relation $f'=-2r\epsilon$ evaluated at $r=a$ gives $B=-2a\epsilon(a)$.
   Multiplying by $da$ and rearranging terms, this relation can be
   written in the form
   \begin{equation}
   \left({B\over 4\pi}\right) d\left( {1\over 4}(2\pi a)\right)= (-T^r_r)(2\pi a)da = (-T^r_r)d(\pi a^2) 
   \end{equation}
   We see that our interpretation carries through in this case as well. Since the 
   temperature is $T = (B/4\pi)$ when $B>0$ and the pressure is 
   $P=-T^r_r$, we can immediately identify the $TdS$ and $PdV$ terms. 
   (The interpretation of the signs proceed as in the $(1+3)$ case and can be provided
   for $B<0$ situation as well.)
    The entropy
   is still  one quarter of the ``area'', $S=(1/4)2\pi a$, and the energy
   vanishes identically: $E=0$.   The vanishing of energy  signifies the fact that at the level of the metric,
   Einstein's equations are vacuous in (1+2) and we have not incorporated any topological
   effects [like deficit angles corresponding to point masses in (1+2) dimensions] in this approach.
   We will see in  section \ref{qt} that the same result can be obtained from the quantum mechanical
   evaluation of the partition function. 

   It would have been nice if the analysis can also be extended to metrics of the form
   in (\ref{basemetric}) with $dL_\perp^2 = dy^2 +dz^2$ and $r$ interpreted in the range
   $-\infty < r < +\infty$ which will include the Rindler spacetime when $f(r)=(1+2gr)$. 
   This, however, cannot be done because both the $TdS$ and $PdV$ terms --- which are
   proportional to the transverse area --- will diverge in this case, making Einstein's equations
   vacuous. As we shall see in section \ref{qt}, one can get around this difficulty in the quantum formulation
    which does not rely on Einstein's equations.

\subsection{ \label{int} Interpretation and Comments}
   
   How does one interpret the differential form of the equation (\ref{keyeqn}) which arises
   from the Einstein's equations written in the form (\ref{albertmod}) ?
   To begin with, we note that this equation is purely local and arises from the classical Einstein's 
   equation connecting $f'$ and $\epsilon$ when $f$ vanishes. (Though I used the trick of
multiplying by $da$ to get (\ref{keyeqn}) one could have got the same result by taking the differential of, say, (\ref{albertmod})
and using Einstein's equations again.)
Therefore any interpretation should
   be purely local. Second, the differential form of the equation shows how the solution changes
   if the location of the  horizon is displaced radially. Endowing the  horizon with
   physical characteristics like entropy, temperature and energy, one can interpret the left
   hand side of (\ref{keyeqn}) as the change in these parameters.
   The right hand side represents the work done by or against the pressure depending on the 
   sign. This relation is independent of how the horizon radius is changed. The standard discussions of first law of black hole dynamics, in which
   a test particle of some mass is made to fall into a black hole, say, is just one way of 
   effecting the change in the horizon size; black hole evaporation could provide another
   process leading to such a change. The formalism does not care what causes the change of 
   the horizon radius. [For a different approach in interpreting
Einstein's equations as a thermodynamic identity see \cite{ted},\cite{tpmpla}.] 

The same result can be stated more formally along these lines: In standard thermodynamics, we can consider two equilibrium states of a system differing infinitesimally in  the extensive variables volume, energy and entropy  by $dV,dE$ and $dS$ while having {\it same} values for the intensive variables temperature ($T$) and pressure ($P$). Then, the first law of thermodynamics asserts that $TdS=PdV + dE$ for these states. In a similar vein, we can consider two spherically symmetric solutions to Einstein's equations with the radius of the horizon differing by $da$ while having the same source $T_{ik}$ and  the same value for $B$.   Then the entropy and energy will be infinitesimally different for these two spacetimes; but the fact that both spacetimes satisfy Einstein's equations shows that $TdS$ and $dE$ will be related to the external source $T_{ik}$ and $da$ by equation (\ref{keyeqn}). Just as in standard thermodynamics, this relation could be interpreted as connecting a sequence of quasi-static equilibrium states. The mathematical description does not distinguish between the `active' interpretation in which the pressure is considered to have done some work against expansion or the `passive' interpretation in which one is comparing the changes in $S,V,E$ of two infinitesimally different states.
[It should be stressed that, in any thermodynamic description of horizons, $P$ does not refer to the pressure of the radiation emitted by the horizon. This is clear from the fact that, in the case of Schwarzschild spacetime with $T_{ik}=0$,
we have $TdS=dE$ though the Hawking radiation will indeed have some pressure.]
   
   In fact, the virtue of the formalism is {\it not} in handling black hole spacetimes
   with single horizon --- for which much more sophisticated, formal and generally covariant
   approach and results are available \cite{BHLS} --- but in its ability to handle spacetimes like
   de Sitter universe, Schwarzschild-de Sitter universe etc. which have horizons
   {\it not} associated with black holes. Compared to  black hole horizons, we have much less
    understanding of  the dynamics of 
    other horizons. (For example, there is no simple analog of the four laws of black hole
    dynamics in the de Sitter spacetime or to the Schwarzschild-de Sitter spacetime.)
   In the local interpretation advocated here, I will consider $E$ to be the energy {\it of the horizon}
   rather then the energy of the spacetime geometry defined asymptotically etc. 
   Thus $E_i, S_i$ and $T_i$ are defined and attributed to each horizon and $P_i dV_i = P_i (4\pi a_i^2) da_i$
   is the work done due to the pressure acting on the transverse proper area of the horizon. The 
   entire description is local
    and allows the formalism to be extended to spacetimes
   with any number of horizons. Virtual displacements of each of the horizon in spherically symmetric
   spacetimes has to satisfy the relationship (\ref{keyeqn}). Each horizon comes with its 
   own temperature, entropy and energy. 
  
   The analysis is  classical except for the crucial periodicity argument which 
   is used to identify the temperature uniquely.  This is again done locally by approximating
the metric by a Rindler metric close to the horizon and identifying the Rindler temperature. (This
idea bypasses the difficulties in defining  and normalizing Killing vectors in spacetimes
which are not asymptotically flat). Without this quantum mechanical
   ingredient one will not be able to fix the constant of proportionality between surface
   gravity and temperature.
   In normal units, equation (\ref{keyeqn}) should read
     \begin{equation}
   {c^4\over G} {B\over 4\pi} d\left( {1\over 4} 4\pi a^2 \right) - {1\over 2}  {c^4\over G}da=
    - T^r_r(a) [4\pi a^2]da 
    \end{equation}
    The second term on the left hand side and the right hand side have the correct
    dimensions for each factor. In the first term on the left hand side, multiplying
    and dividing by $\hbar$ allows us to write
    \begin{equation}
    {c^4\over G} {B\over 4\pi} d\left( {1\over 4} 4\pi a^2 \right) =
    {c^3\over G\hbar} {B\hbar c\over 4\pi } d\left( {1\over 4} 4\pi a^2 \right) 
    =T d({A\over 4L_P^2})
    \end{equation}
    with $T=(B\hbar c/4\pi)$ being the temperature in energy units.
    It is obvious that quantum mechanics enters only through the factor 
    $\hbar$ which appears in expressing the temperature in terms of the
    surface gravity. Of course, one needs $\hbar$ along with 
   $G$ and $c$  to obtain a quantity with dimensions of area.

    One can extend much of this analysis to a slightly more general
    case in which $g_{00} = f_1(r)$ and $g_{11} = -1/f_2(r)$ with $f_1 \ne f_2$ in a fairly 
    straightforward manner.
    Using the relevant Einstein's equations for this case, simple algebra shows that:
    (i) At the horizon,
    both $f_1   $ and $f_2$ will vanish,
    (ii) $T^t_t = T^r_r$ at the horizon, though in general they are not equal and
    (iii) equation (\ref{albertmod}) is satisfied. 
    To prove (i) let us assume that  $f_2(r)=0$ at some $r=a$ for the first time as we approach 
    from the right and $f_1(r)>0$ at $r>a$.
    If $f_1(r)>0$ at $r=a$, then $t$ direction will 
    remain time like and $(\theta,\phi)$ directions will be space like at $r=a$. 
    But since the horizon at $r=a$ must be a null surface, it must have a null tangent direction and symmetry
    requires this to be the $t$ direction. Hence $f_1$ must vanish at $r=a$. Using the behaviour of Einstein's
    equations it is also possible to prove the converse: that is, if  $f_1(r)=0$ at $r=a$, then $f_2(r)=0$ at $r=a$.
    The result (ii) can be proved by expanding $f_1(r),f_2(r)$ around $r=a$ with derivatives $B_1,B_2$
      at
    $r=a$ and substituting into the expression for the scalar curvature. 
    The scalar curvature will be finite at $r=a$ only if  $T^t_t = T^r_r$ at the horizon. 
    Using (ii) in the Einstein's equations will lead to (iii).
   Thus the only non trivial assumption used
    in the above analysis is that of spherical symmetry.

\subsection{Dimensionality of horizons}    
    
    Incidentally, this analysis throws light on another issue, viz., the effective dimension
    of the horizon. Bekenstein and Mayo have argued  \cite{bekoned}  
 that the  horizon of the Schwarzschild black hole is effectively one dimensional.
     The argument
    relies on the fact that the rate of flow, $\dot S$, of entropy or information in a channel
    with power ${\cal P}$ scales as $\dot S\propto {\cal P}^{1/2}$ for an one 
    dimensional channel, independent of other details. For higher dimensional channels
    it is not possible to obtain the unique index of $(1/2)$ in the above
    relation. In the case of  a blackbody radiation emerging from a 
    surface of area $A$ and temperature $T$, we have $\dot S \propto AT^3$
    while ${\cal P}\propto AT^4$. In general, there is no relation between
    $A$ and $T$ and one cannot relate $\dot S$ to ${\cal P}$.
    In the case of spherically symmetric spacetimes with horizons considered
    above, $A\propto a^2$ and $T ^3\propto |f'(a)|^3\propto B^3$. Taking $\dot S\propto
    B^3 a^2$,  ${\cal P}\propto B^4 a^2$ and using (\ref{albertmod}) it follows that      
    \begin{equation}
    {\dot S\over \sqrt{\cal P}} \propto |B|a = |1- a^2 \epsilon(a)|
    \label{jacob}
    \end{equation}
    Therefore Bekenstein's result can be generalized to any spacetime
    in which $\epsilon (a)$ vanishes or is proportional  to $a^{-2}$. The de Sitter universe
    is an example of the latter case for which $\epsilon (a) a^2 = 3$ allowing us to 
    conclude that pure de Sitter horizon is also of effective one dimension.
    Equation (\ref{jacob})  shows that the result does {\it not} extend to
    other cases like, for example, the Schwarzschild-de Sitter spacetime
    or even the Reissner Nordstrom black hole. 
       The reason for this failure is related to  a feature which was mentioned earlier.
     Spacetimes like Schwarzschild or de Sitter have only one free parameter
    in the metric (like $M$ or $H^{-1}$) and hence the scaling of all other thermodynamic 
    parameters is uniquely fixed by purely dimensional considerations. This is the reason
    why combination like $Ba$ are pure numbers in these cases. But, in general, the 
    area of the horizon  (and hence the entropy) is determined by the location of the zero of the function $f(r)$
    while the temperature --- determined by periodicity considerations ---
    is determined by the value of $f'(r)$ at the zero. For a general function, of course,
    there will be no relation between the location of the zero and the slope of the 
    function at that point. Thus, we cannot expect $\dot S/\sqrt{\cal P}$
    to have a simple form in the general case.

    \section{ \label{qt} Partition function in quantum theory}
    
    The analysis given above provided expressions for $S$ and $E$ using Einstein's equation. I
    will now show that the same result can be obtained in quantum theory {\it without using 
    Einstein's equations}, for a slightly wider class of metrics \cite{tpprl}.
    
    A wider class of  spacetimes, analyzed in the literature, has the form in (\ref{basemetric})
    where $f(r)$ vanishes at some surface $r=a$, say, with $f'(a)\equiv B$ remaining finite
    and $dL_\perp^2$ interpreted more generally.
     When $dL_\perp^2$
  is taken as the metric on 2-sphere and $r$ is interpreted as the radial coordinate $[0\leq r\leq \infty]$, equation (\ref{basemetric}) covers a variety of spherically symmetric spacetimes  (including Schwarzschild, Reissner-Nordstrom, de Sitter etc.) with a compact horizon at $r=a$. [This is the case discussed earlier. I do not assume that there is only one horizon; symbols like $a,B ...$ etc refer to different horizons if more than one horizon is present and should be thought of
  as $B_i,a_i,....$ with $i=1,2,....$. For simplicity of notation, I suppress the subscript $i$ in what follows.] If $r$ is interpreted as one of the Cartesian coordinates $x$ with $(-\infty\leq x\leq \infty)$ and $dL_\perp^2=dy^2+dz^2, f(x)=1+2gx,$ equation (\ref{basemetric}) can describe the Rindler frame in flat spacetimes. 
   Finally $0\le r\le \infty$ with $dL_\perp^2 = r^2 d\theta^2$ will describe circularly symmetric $(2+1)$ dimensional spacetimes.
  We shall first concentrate on compact horizons in $(3+1)$  with $r$ interpreted as radial coordinate,
and comment on the other cases at the end.

 Since the metric is static, Euclidean continuation is trivially effected by $t\to
  \tau=it$ and an examination of the conical singularity near $r=a$ [where $f(r) \approx B(r-a)$] shows that $\tau$ should be interpreted as periodic with period $\beta=4\pi/|B|$ corresponding to the temperature $T=|B|/4\pi$.  Let us consider any one of the horizons with the temperature $T=|B|/4\pi$.
 The class of metrics in
  (\ref{basemetric}) with the behaviour $[f(a)=0,f'(a)=B]$ constitute a canonical ensemble at constant temperature  since they all have the same temperature $T=|B|/4\pi$ . The partition function for this ensemble is given  by the path integral sum
   \begin{eqnarray}
    Z(\beta)&=&\sum_{g\epsilon {\cal S}}\exp (-A_E(g))  \\
 &=&\sum_{g\epsilon {\cal S}}\exp \left(-{1\over 16\pi}\int_0^\beta  d\tau \int d^3x \sqrt{g_E}R_E[f(r)]\right)\nonumber
     \label{zdef}
     \end{eqnarray}
  where I have made the Euclidean continuation of the Einstein action and imposed the periodicity in $\tau$
  with period $\beta=4\pi/|B|$. [For static spacetimes, there is some ambiguity regarding the overall
  sign when the Euclidean continuation is performed. The convention adopted here is  as follows:  First note that positive definite Euclidean metric requires one to work with a signature $(-+++)$ while
  I am using $(+---)$. Changing the signature in a diagonal metric is equivalent to changing the signs of components of metric tensor: $g_{ik}\to -g_{ik}$. This leads to $\Gamma_{i,kl}\to -\Gamma_{i,kl}; 
  \Gamma^i_{kl}\to \Gamma^i_{kl}; R^i_{\phantom{i} jkl}\to R^i_{\phantom{i} jkl}; R_{jl}\to R_{jl}; R\to -R$. The
  Euclidean continuation is  effected through $it=\tau$ in a spacetime with signature $(-+++)$.]
  The sum in (\ref{zdef}) is restricted to the set ${\cal S}$ of all metrics of the form in 
  (\ref{basemetric}) with the behaviour $[f(a)=0,f'(a)=B]$ and the Euclidean Lagrangian is a functional of $f(r)$.
  No source term or cosmological constant (which cannot be distinguished from certain form of source) is included since the idea is to obtain a result which depends purely on the geometry.
   The spatial integration will be restricted to a region bounded by the 2-spheres $r=a$ and $r=b$, where
  the choice of $b$ is arbitrary except for the requirement that  within the region of integration the Lorentzian
  metric must have the proper signature with $t$ being a time coordinate.  Using the result
  \begin{equation}
  R={1\over r^2}{d\over dr}(r^2 f') -{2\over r^2}{d\over dr}\left[r(1-f)\right]\label{reqn}
  \end{equation}
  valid for metrics of the form in (\ref{basemetric}) [with signature (+ - - -)], a
   straight forward calculation shows that
   \begin{eqnarray}
  - A_E&=&{\beta\over 4}\int_a^b dr\left[-[r^2f']'+2[r(1-f)]'\right] \nonumber\\
&=&{\beta\over 4}[a^2B -2a]+Q[f(b),f'(b)]
     \label{zres}
     \end{eqnarray}
where $Q$ depends on the behaviour of the metric near $r=b$ and we have used the 
conditions $[f(a)=0,f'(a)=B]$. The sum in (\ref{zdef}) now reduces to summing over the values of $[f(b),f'(b)]$
with a suitable (but unknown) measure. This sum, however, will only lead to a factor which we can ignore in deciding about the dependence of $Z(\beta)$ on the form of the metric near $r=a$. Using $\beta=4\pi/B$
(and taking $B>0$, for the moment)  the final result can be written in a very suggestive form:
 \begin{equation}
  Z(\beta)=Z_0\exp \left[{1\over 4}(4\pi a^2) -\beta({a\over 2})  \right]\propto 
  \exp \left[S(a) -\beta E(a)  \right]
     \label{zresone}
     \end{equation}
with the identifications for the entropy and energy being given by:
\begin{equation}
S={1\over 4} (4\pi a^2) = {1\over 4} A_{\rm horizon}; \quad E = {1\over 2} a = \left( {A_{\rm horizon}\over 16 \pi}\right)^{1/2}
\end{equation}

  This is exactly the result obtained earlier except that  we have {\it not} used the Einstein's equations.
   Instead, the partition function was evaluated with two very natural conditions: $f(a) =0$ making 
    the surface $r=a$ a compact horizon and $f'(a) = $ constant which is the 
    proper characterization of the canonical ensemble of spacetime metrics. Since temperature
    is well defined for the class of metrics which I have considered, this canonical ensemble is defined without any   ambiguity.  This allows me to  sum over a  class of spherically symmetric spacetimes at one go rather than
    deal with, say, black hole spacetimes and de Sitter spacetime separately. 
    
     The idea of using  a canonical ensemble is not new and one of the earliest attempts is
  by \cite{gibhawk}; similar ideas have been explored by different people \cite{actionpap} mostly in the context of BH spacetime.
   What is new  are the following features: (a) I have not added any boundary terms to the
  action  and in fact $R\sqrt{-g}$ in (\ref{reqn})  is a total derivative.
 A detailed description of action functionals are given in 
   \ref{appb} showing that the same result is obtained from the surface terms, if done correctly.
  (b) I have not invoked WKB limit to evaluate the action using a solution to Einstein's equations. 
  Normally, the path integral in (\ref{zdef}) is formidable to calculate and is ill defined without the boundary terms. I can manage with (a) and (b) above only because I confine the sum to metrics in the set ${\cal S}$.
   Conceptually,
a canonical ensemble for a minisuperspace of metrics  of the form in (\ref{basemetric})  should 
be constructed by keeping the temperature constant {\it without} assuming the metrics
to be the solutions of Einstein's equation; this is what I do and exploit the form of $R$ given by (\ref{reqn}).
Since this action involves second derivatives, it is not only allowed but even required to fix both $f$ and $f'$
at the boundaries. 
Moreover, since I am summing over an arbitrary class of $f(r)$, the behaviour of the function
$f(r)$ at $r=b$ is not in any way related to the parameter $a$ or parameter $B$
which enters the analysis at the horizon. In the conventional
analysis $f$ is chosen to be a classical solution and hence its value at any radius is 
predetermined in terms of the parameters of the solution (like $M$ in the Schwarzschild metric).

 The temperature is determined by using Euclidean periodicity condition
near the horizon and is related to the derivative of $f$ by $T=|f'(a)|/4\pi$. 
In the literature one sometimes uses the locally defined Tolman temperature $T_{\rm local}(x)
 = T_R(g_{00}(x_R)/g_{00}(x))^{1/2}$ where $x_R$ is some reference point.
This local temperature will diverge at the horizon if $T_R$ is defined as  a finite quantity elsewhere.
In the approach taken here, all these issues are by passed by directly defining the temperature
by expanding $f(r)$ in a Taylor series near the horizon and using the locally defined Rindler 
temperature. It is important that one uses this quantity because there is no well defined notion of
temperature in spacetimes with multiple horizons (see \ref{appa}).

We get the same result as as obtained earlier in the case of horizons with $B<0$ like the
     de Sitter universe as well. In this case, $f(r) = (1-H^2r^2)$, $a=H^{-1}, B=-2H$.
Since the region where $t$ is time like is ``inside'' the horizon, the integral for $A_E$ in (\ref{zres}) should be taken from some arbitrary value $r=b$ (which could even be the origin) to the horizon at $r=a$ with $a>b$. So the horizon contributes in the upper limit of the integral
introducing a change of sign in (\ref{zres}). Further, since $B<0$, there is another negative sign in the area term
from $\beta B\propto B/|B|$. Taking all these into account we get, in this case, 
\begin{equation}
  Z(\beta)=Z_0\exp \left[{1\over 4}(4\pi a^2) +\beta({a\over 2})  \right]\propto 
  \exp \left[S(a) -\beta E(a)  \right]
     \label{zrestwo}
     \end{equation}
giving
$S=(1/ 4) (4\pi a^2) = (1/ 4) A_{\rm horizon}$ and  $E=-(1/2)H^{-1}$ as before. 

Let us next consider the (1+2) dimensional spacetime.
 In $D=(1+2)$,  
 metrics of the type in (\ref{basemetric}) with $dL^2_\perp=r^2 d\theta^2$ will
 have the scalar curvature 
\begin{equation}
R={1\over r}{d\over dr}(rf') + {f'\over r} 
\end{equation}
The action will now become
 \begin{equation}
 -A_E = {B\over 16\pi} \int_a^b 2\pi dr [-(rf')' -f'] = {1\over 4} (2\pi a) + Q[f'(b)]
 \end{equation}
 leading to $Z= Z_0 \exp [ (1/4) A_{2D}]$  where the ``area'' of the 2D horizon is $2\pi a$. This will  
  give $S=(1/4)(2\pi a)=(1/4)A_{\rm horizon}$ with $E=0$. The vanishing of energy  signifies the fact that at the level of the metric, Einstein's equations are vacuous in (1+2) and we have not incorporated any topological
effects [like deficit angles corresponding to point masses in (1+2) dimensions] in our approach.

Interestingly enough, this formalism can also handle spacetimes like Rindler unlike the approach 
discussed in the last section
which could only work for compact transverse areas.
For the spacetimes with planar symmetry for which (\ref{basemetric}) is still applicable with $r=x$ being a Cartesian coordinate and $dL_\perp^2=dy^2+dz^2$. In this case $R=f''(x)$ and the action becomes
\begin{eqnarray}
-A_E&=&{1\over 16\pi}\int_0^\beta d\tau\int dy dz \int_b^a dx f''(x)\nonumber\\
&=& {\beta\over 16\pi} A_\perp f'(a)+Q[f'(b)]
\label{rindleraction}
\end{eqnarray}
where we have confined the transverse integrations to a surface of area $A_\perp$. If we now sum over
all the metrics with $f(a)=0,f'(a)=B$ and $f'(b)$ arbitrary, the partition function will become
\begin{equation}
Z(\beta)=Z_0\exp({1\over 4}A_\perp)
\end{equation}
which shows that planar horizons have an entropy of (1/4) per unit transverse area but zero energy. This includes
Rindler frame as a special case. Note that if we freeze  $f$ to its Rindler form $f=1+2gx$, 
(by demanding the validity of Einstein's equations in the WKB approach, say)
then $R=f''=0$ as it should.
In the action in (\ref{rindleraction}), $f'(a)-f'(b)$ will give zero. It is only because I am
 {\it not} doing a WKB analysis --- but 
 varying $f'(b)$ with fixed $f'(a)$ --- that I obtain an entropy for these spacetimes.

It is interesting that  this approach, which does not use  classical Einstein's equation, gives the same result as the one obtained earlier using  Einstein's equation. One  difference is that the partition function allows us to identify $S$ and $E$ without recourse to $P$ or $V$ while the in the earlier approach we essentially rewrote
Einstein's equation as first law of thermodynamics (which included the $PdV$ term) by multiplying by $da$. 
The fact that the approach based on the action functional leads to the thermodynamic interpretation directly [without the use of Einstein's equation] suggests that there could be a deep connection between the form of the Einstein action and the thermodynamic interpretation of spacetimes with horizons. This is indeed true and some aspects of this connection is explored in detail elsewhere \cite{tpmpla} \cite{tpprl}. In particular, it was shown in \cite{tpmpla} that the Einstein-Hilbert action can be expressed as the ``free energy of
spacetime"  and  Einstein's equations may be interpreted as describing the {\it thermodynamics of spacetime} while quantum gravity is needed to describe the statistical mechanics of the microscopic degrees of freedom
making up the spacetime. 

Both the classical and quantum descriptions developed here
 used spherical symmetry; in the first approach, we used Einstein's equations for the spherically symmetric spacetime while in the second approach we used the form of $R$ valid for spherically symmetric spacetimes.  One may wonder whether this assumption is essential or not. Given the broader interpretation of Einstein-Hilbert action as a free energy of spacetime,   it seems likely that one will be able to generalise the results of this paper to any horizon
without having to assume spherical or planar symmetry. This issue is under investigation. 

I thank Apoorva Patel, S.Shankaranarayanan and Suneeta Varadarajan for useful comments.

\appendix

\section{\label{appa}Spacetimes with multiple horizons}

Metrics in the form in (\ref{basemetric}) with $f(r)$ having a simple zero at one or more points,
 $r=a_i , i=1,2,3, ...$, exhibit coordinate singularities at $r=a_i$. The coordinate $t$ alternates
 between being time like and space like when each of these horizons are crossed.
 Since all curvature invariants are well  behaved at the horizons, it will be possible to introduce
 coordinate patches such that the metric is also well behaved at the horizon. There is a 
 general procedure for doing this in the case of spacetimes with a single horizon,
  which can be used to handle multi horizon spacetimes as well.

Concentrating on the $(t,r)$ 
plane, we first introduce 
the tortoise coordinate $\xi$ so that the metric becomes
\begin{equation}
ds^2 = f(r) dt^2 - {dr^2\over f(r)}=f(\xi)(dt^2 - d\xi^2);\quad \xi = \int {dr\over f(r)}
\end{equation}
Switching over to the light cone coordinates $(u,v)$  by
\begin{equation}
 u= t-\xi; \qquad v=t+\xi
\end{equation}
the metric becomes
\begin{equation}
  ds^2 = f(\xi)(dt^2 - d\xi^2) = f[{1\over 2}(v-u)]du dv
\end{equation}
The coordinate systems $(t,\xi), (u,v)$  are still singular near the horizons $r\approx a_i$. Assuming
$f(r) \approx B_i(r - a_i)$ near $r=a_i$ [with $B_i=f'(a_i)$], it is easy to see that near the horizon
the line element has the form 
\begin{equation}
 ds^2\approx B_ia_i e^{(B_i/2)(v-u)} dv du
 \label{cone}
 \end{equation}
 which is badly behaved because $(v-u)\propto \xi$ has a logarithmic singularity near
 $r=a_i$. However, the form of the metric in (\ref{cone}) suggests 
a natural coordinate transformation from 
$(u,v)$ to a non singular coordinate system $(U,V)$ with 
\begin{equation}
 V_i = (2/B_i) \exp[(B_i/2)v]; \quad U_i=-(2/ B_i) \exp[(-B_i/2)u]
 \label{nonsing}
 \end{equation}
in terms of which the metric becomes $ds^2 = -(4f/ B_i^2 U_iV_i)$ $ dU_idV_i$.
By construction, the $(U_i,V_i)$ coordinate system is regular on the horizon at
$r=a_i$, since the combination $(f/U_iV_i)$ is finite
on the horizon.
The transformation (\ref{nonsing}) will lead to the familiar Kruskal type coordinate system in all  cases
with a single horizon like the
 Schwarzschild, Reissner Nordstrom or de Sitter  manifold. In these cases $i=1$ and we need only
 one coordinate  patch with $(U,V)$ to cover the manifold. 
 
 When there is more than one
 horizon,  we need to introduce one Kruskal like coordinate patch
 for each of the horizons; the $(u,v)$ coordinate system is unique in the manifold but
 the $(U_i,V_i)$ coordinate systems are different for each of the horizons since the 
 transformation in (\ref{nonsing}) depends explicitly on $B_i$'s which are (in general)
 different for each of the horizons. In such a case,
 there will be regions of the manifold in which more than one Kruskal like patch
 can be introduced. The compatibility between these coordinates  can provide interesting constraints.

    It is easy to develop the quantum field theory in the $t-r$ plane if we treat it 
    as a $(1+1)$ dimensional spacetime. In this case, the solutions to the wave equations
    are just arbitrary functions of the null coordinates in a conformally flat coordinate system.
    Since $(u,v)$ as well as all the $(U_i, V_i)$ coordinates retain the conformally flat
    nature of the $(1+1)$ dimension, we can define suitable mode functions and 
    vacuum state in a straightforward manner.    
    The outgoing and ingoing modes of the kind 
$(4\pi \omega)^{-1/2} $ $[\exp(-i\omega u), \exp(-i\omega v)]$
defines a static vacuum state. This is called Boulware vacuum in the 
case of Schwarzschild black hole, but it can be defined in a general spacetime
of the kind we are studying (see reference \cite{tprealms}).
The modes of the kind 
$(4\pi \omega)^{-1/2} [\exp(-i\omega U_i), $ $\exp(-i\omega V_i)]$
define another  vacuum state called Hartle-Hawking vacuum in the 
case of Schwarzschild black hole. (The modes of the kind
$(4\pi \omega)^{-1/2}  [\exp(-i\omega U), $ $\exp(-i\omega v)]$
 define the analogue of Unruh vacuum which, however, we will not need.)
   The transformations between the Boulware and Hartle-Hawking 
    modes are identical to those in the case of Schwarzschild black hole with
    the surface gravity at the corresponding horizon occurring in the 
    transformations. The study  of either Bogoliubov coefficients or the stress-tensor
   expectation values will now show that the Hartle-Hawking vacuum has an equilibrium
   temperature $T_i =|B_i|/4\pi$.  The Bogoliubov coefficients between two different Hartle-Hawking
   modes, of course, have no simple interpretation, showing that there is a conceptual problem in 
   defining a unique temperature in the overlapping region. 
   
   This difficulty is apparent even from an examination of the coordinate 
   transformations. 
    Consider, for example, the region between two consecutive horizons $r_n <r<r_{n+1}$ in which $t$ is time like. The coordinates
    $(U_i, V_i)$ with $i=n, n+1$ overlaps in this region.
    Euclideanisation of the metric can be easily effected in the region
    $r_n<r<r_{n+1}$ by taking $\tau =it$. This will lead to the transformations 
    \begin{eqnarray}
    U_{n+1} = -U_n \exp [(B_{n+1} + B_n)((-i\tau -\xi)/2)];&&\nonumber \\
     V_{n+1} = -V_n \exp[ -(B_{n+1} + B_n)((-i\tau+\xi)/2)]&&
    \end{eqnarray}
    Obviously, single valuedness can be maintained only if the period of $\tau$ is an integer multiple
    of $4\pi / (B_{n+1}+B_n)$. 
    More importantly, we get    from (\ref{nonsing}) the relation 
    \begin{equation}
    U_i +  V_i  = {4\over B_i} \exp \left({B_i \xi\over 2}\right) \sinh \left(-i{B_i \tau\over 2}\right)
    \end{equation}
    which shows that $(U_i, V_i)$ can be used to define values of $\tau$ only up to integer multiples of
    $4\pi / B_i $  in each patch. But since $(U_n, V_n)$ and $(U_{n+1}, V_{n+1})$ are to be 
    well defined coordinates in the overlap, the periodicity $\tau \to \tau +\beta$ which 
    leaves both the sets $(U_n, V_n)$ and $(U_{n+1}, V_{n+1})$
     invariant must be such that $\beta $ is an integer
    multiple of both $4\pi / B_n $ and $4\pi / B_{n+1}$. 
    This will require $\beta = 4\pi n_i/B_i$ for all $i$ with $n_i$ being a set of integers. This,
    in turn, implies that $B_i/B_j =n_i/n_j$ making the ratio between any two surface 
    gravities a rational number. Since there is no physical basis for such a condition,
    it seems reasonable to conclude that these difficulties arise
    because of our demanding the existence of a finite periodicity $\beta$ in the
    Euclidean time coordinate. This demand is related to
    an expectation of thermal equilibrium which is violated in spacetimes with 
    multiple horizons having different temperatures. Hence, such spacetimes
    will not have a global notion of temperature, entropy etc. It seems,
     however, unlikely that we cannot attribute a temperature to a black
     hole formed in some region of the universe just because the universe
     at the largest scales is described by a de Sitter spacetime, say.
     One is led to searching for a local description of the dynamics of all
     horizons. 
    
 \section{\label{appb}The choice of action functional}

 Action functionals in classical theory are essentially tools to obtain
   equations of motion and provide a book keeping of the symmetries of the 
   theory. The numerical value of the action does not play a significant role in 
   classical theory. The situation is different in quantum theory in which the actual
   value of the action may be of relevance in evaluating a path integral
   or in interpreting a partition function. 
   
   In the case of general relativity, the generally covariant action is the Einstein-Hilbert
   action $A_{\rm EH}$ given by 
 \begin{equation}
   A_{\rm EH} \equiv {1\over 16\pi} \int R\sqrt{-g} d^4x 
   \end{equation}  
    Since $R\sqrt{-g}$ is linear in second derivatives of the metric tensor,
    it is possible to write  this action in the form
    \begin{eqnarray}
   A_{\rm EH}  &=&{1\over 16\pi}\int L_{\rm SD} \sqrt{-g} d^4x  - {1\over 16\pi}\int \partial_c P^c  d^4x \nonumber \\
  &\equiv& A_{\rm SD} - {1\over 16\pi}\int \partial_c P^c  d^4x
  \label{defasd}
   \end{eqnarray}   
    where 
    \begin{equation}
    L_{\rm SD} = g^{ab} \left( \Gamma^i_{ja} \Gamma^j_{ib} -  \Gamma^i_{ab} \Gamma^j_{ij}\right)
    \end{equation}
    \begin{equation}
    P^c = \sqrt{-g} \left( g^{ck} \Gamma^m_{km} - g^{ik} \Gamma^c_{ik}\right) =
    2g^{cb} \partial_b \sqrt{-g} + \sqrt{-g} \partial_b g^{bc}
    \label{defpc}
     \end{equation}
    Equation (\ref{defasd}) defines the Schroedinger-Dirac action $A_{\rm SD}$ which is quadratic 
    in the Christoffel symbols and does not contain second derivatives of the metric 
    tensor. 
    
    It is possible to write the second term 
    in (\ref{defasd}) in a different manner which is often used in the literature.
If we foliate the spacetime by a series
of spacelike hyper-surfaces  ${\cal S}$ with $n^i$ as normal, then $g^{ik}=h^{ik}+n^in^k$ where $h^{ik}$ is the induced metric on ${\cal S}$. 
It is  conventional to define
    a quantity called extrinsic curvature $K_{ab}$ such that 
    \begin{equation}
    K_{ab} = \nabla_a n_b + n_a a_b
    \end{equation}
    where $a^b \equiv n^c\nabla_c n^b$ is the ``acceleration'' corresponding to 
    trajectories with ``four velocity'' $n^a$. (This makes sense on the $x^0 = $ constant
    space-like surfaces since the unit normal $n^a $ to such surfaces can be
    thought of as the four velocities tangential to ${\bf x} =$ constant curves.
   On the surfaces like $x^1 = $ constant, say, the normal will not be
   time like and $a^b$ should be just treated as a vector normal to $n^b$ and tangential
   to the surface element. It does not correspond to 
   a physical acceleration.)
   In all the surfaces, $n^b a_b =0$ allowing us to write $K = \nabla_a n^a$.
Given the covariant derivative $\nabla_i n_j$ of the normals to ${\cal S}$, one can construct only  three vectors
$(n^j\nabla_j n^i, n^j\nabla^i n_j, n^i\nabla^j n_j)$ which are linear in covariant derivative operator. The first one is
the acceleration $a^i=n^j\nabla_j n^i$; the second identically vanishes since $n^j$ has unit norm; the third,
$u^i K$, is
proportional to the trace of the extrinsic curvature $K=\nabla^j n_j$ of ${\cal S}$. Hence, the the second term 
    in (\ref{defasd}) --- which is linear in the derivatives of the Christoffel symbols --- must be the four divergence
of a linear combination of $u^i K$ and $a^i$. So the corresponding
term in the action must have the form 
\begin{equation}
A_{\rm surface} = \int d^4 x \, \sqrt{-g} \nabla_i \left[ \lambda_1 K u^i + \lambda_2 a^i\right]
\label{threedaction}
\end{equation}
where $\lambda_1$ and $\lambda_2$ are numerical constants. This final result --- though
 probably not the 
reasoning given above --- is known in the conventional
$(3+1)$ formalism  (see e.g., equation (21.88) of \cite{irrmass}).
Since $u^ia_i=0$, the spacelike boundaries at $x^0=$ constant gets contribution only
from $K$ while the time like surfaces like $x^1=$ constant gets contribution from the normal component
of the acceleration $a^i\hat n_i$ where $\hat n_i$ is the normal to the time like surface. The numerical constants can be easily evaluated by choosing some simple metric, thereby fixing the surface term.

 Such a detailed analysis shows that the surface integral can actually be written as
    \begin{equation}
     {1\over 16\pi}\int_{\cal V} \partial_c P^c  d^4x = \sum{1\over 8\pi} \int_{\partial {\cal V}} d^3x\, \sqrt{h}\, K
     \label{pctok}
     \end{equation}
     It is therefore possible to define the $A_{\rm SD}$ by either of the relations
       \begin{eqnarray}
   A_{\rm SD}  &=&A_{\rm EH} + {1\over 16\pi}\int \partial_c P^c  d^4x\nonumber \\
  &\equiv& A_{\rm EH} + \sum {1\over 8\pi} \int_{\partial {\cal V}} d^3x\, \sqrt{h}\, K
  \label{final}
   \end{eqnarray}   
   The following points need to be noted regarding this result:
(a) From the origin of this equation we know that neither the left  hand side
   nor the second term in the right hand side is generally covariant.
  (b) The action $A_{\rm EH}$ vanishes in flat spacetime. The action $A_{\rm SD}$ 
   does not, in general, vanish in flat spacetime if non Cartesian coordinates
   are used.  
   (c) The numerical values of all these actions can be formally divergent
   if integration domain is unbounded either in space or time. In the case of 
   $A_{\rm SD} $  this can happen even in flat spacetime
   if curvilinear coordinates are used.

      The action $A_{\rm EH}$ vanishes for any vacuum solution of Einstein's
   equation and in particular for the Schwarzschild metric. If one insists on
   using WKB value of action to interpret spacetime thermodynamics, then
   $A_{\rm EH}$ will not help even in the simplest case of Schwarzschild black hole.
   Motivated by this, most of the previous workers in this field \cite{actionpap}
   have used $A_{\rm SD}$ in order to obtain a non zero
   value for the action for the Schwarzschild black hole. In fact, it is often done by using
the second expression  (involving the surface integral).    
 While this is done fairly routinely in the literature, it is easy to go wrong unless
  some caution is exercised, especially in deciding
  which are the surfaces over which one must do the summation.
 
 To illustrate this, I shall consider evaluation of the action for the 
  class of metrics studied in this paper --- which have been repeatedly analyzed in the literature,
  though wrongly at times. The line element  is
   \begin{equation}
     ds^2=f(r)dt^2-f(r)^{-1}dr^2 - r^2(d\theta^2+\sin^2\theta
     d\phi^2)
     \label{mymetric}
     \end{equation}
      I choose the four dimensional region ${\cal V}$ to be bounded by two spacelike surfaces ( $S_0,S_\beta$)
      at $t=0,t=\beta$  and two time like surfaces  ($R_a,R_b$) at $r=a,r=b$. Since the metric is static, the time derivative in $\partial_c P^c$
      does not contribute (or, rather  the contributions from $S_0$ and $S_\beta$ cancel each other).
      We thus only need to do the integral over $R_a,R_b$ and --- in these integrals ---
       the integration over time reduces to multiplication by $\beta$. Thus
     \begin{equation}
      {1\over 16\pi} \int_{\cal V} \partial_c P^c  d^4x= {\beta\over 16\pi}\int \partial_\alpha P^\alpha  d^3x;
       \qquad (\alpha=1,2,3)
      \end{equation}
      If we convert this into a surface integral of the radial component $P^r$ 
      over the bounding surfaces $R_a,R_b$, then
      we will get
      \begin{eqnarray}
      {1\over 16\pi} \int_{\cal V} d^4x\, \partial_c P^c &=& {\beta\over 16\pi} \int d^3 x 
      \, \partial_\alpha P^\alpha\nonumber \\
      &=& {\beta\over 16 \pi } \int_{\theta=0}^{\theta=\pi} d\theta \int_0^{2\pi} d\phi\  P^r\bigg|_a^b
       \label{keyone}
       \end{eqnarray}
      where the vertical line denotes that the integral should be evaluated at $r=b$ and $r=a$
      and the latter should be subtracted from the former. The relevant component is
      \begin{equation}
      P^r = 2 g^{rr} \partial_r \sqrt{-g} + \sqrt{-g} \partial_r g^{rr} = - r^2 \sin\theta \left( {4f\over r} + f'\right)
      \end{equation}
      (note that the $\sqrt{-g}$ factors are built into $P^c$; the $d^nx$ just stands for $d^0x....d^{n-1}x$)
      giving the result
      \begin{equation}
      {1\over 16\pi} \int d^4x\, \partial_c P^c = {\beta\over 4} \left[ - r^2 f' - 4 rf\right]^b_a
      \label{wrongone}
      \end{equation}
      This exactly what we would have got by integrating $2K=2\nabla_a n^a$ over the $r=$constant
      surface. Using 
      \begin{equation}
      K = {1\over \sqrt{-g}}\partial_a (\sqrt{-g}g^{ab} n_b); \quad n_a= (0,1,0,0) {1\over \sqrt{f}}
      \end{equation}
      we get  the extrinsic curvature to be
   \begin{equation}
   K = - \sqrt{f} \left( {1\over 2} {f'\over f} + {2\over r}\right) 
   \end{equation}
   The integral on the right hand side of  (\ref{pctok}) for the $r=$constant surface is
   \begin{eqnarray}
    {1\over 8\pi} \int_0^\beta dt &&\hskip-2em \int_0^{\pi} d\theta  \int_0^{2\pi} d\phi \, \sqrt{|h|} K ={\beta\over 8\pi}
     \int 2\pi d\theta\, \left( \sqrt{f} r^2 \sin\theta\right) K \nonumber\\
     &=& {\beta\over 2} r^2 \sqrt{f} \left[ - \sqrt{f} \left( {1\over 2} {f'\over f} + {2\over r}\right) \right]
     = {\beta \over 4} \left[ - r^2 f' - 4 rf\right] 
     \label{wrongtwo}
   \end{eqnarray}
   This result has been obtained in the literature several times, especially in the context of Schwarzschild
   spacetime in which $R\sqrt{-g}$ vanishes and the only contribution to $A_{SD}$ is from
   this term. The term in the right hand side will contribute $-4\pi M^2=-(1/4)$(Horizon Area)
   when evaluated on the horizon $r=2M$ and is often related to the entropy.
   
   In spite of the simplicity of the calculation, {\it this result is algebraically wrong}.
   
   Proving it  wrong is quite trivial and can be done by computing (\ref{pctok}) by evaluating the left hand side
   directly. In $\partial_c P^c$, derivatives with respect to $t$ and $\phi$ vanish, giving
   \begin{equation}
   \partial_c P^c = {\partial P^r\over \partial r} + {\partial P^\theta \over \partial \theta}
   \end{equation}
   Integrating over the four volume the first term will give
   \begin{eqnarray}
  &&{1\over 16\pi} \int_0^\beta dt\int_0^{\pi} d\theta  \int_0^{2\pi} d\phi \int_a^b dr\,\left( {\partial P^r\over 
   \partial r} \right)\nonumber \\
   && \hskip 5em = {1\over 16\pi} \int_0^\beta dt \int_0^{\pi} d\theta \int_0^{2\pi} d\phi \, P^r \bigg|_a^b 
   \end{eqnarray}
   which is precisely the contribution in (\ref{keyone}). This will lead to the result quoted above. But
   there is a contribution from the second term! Since
   \begin{equation}
   P^\theta = 2 g^{\theta\theta} \partial_\theta \sqrt{-g} + \sqrt{-g} \partial_\theta g^{\theta\theta}  
   =-{2\over r^2} r^2 \cos\theta = - 2 \cos\theta
   \end{equation}
   the second term gives
   \begin{eqnarray}
&&{1\over 16\pi} \int_0^\beta dt  \int_0^{2\pi} d\phi \int_a^b dr\int_0^{\pi} d\theta\, {\partial P^\theta\over 
   \partial \theta} \nonumber \\
   &&\quad= {\beta\over 8} \int_a^b dr\, \left[ P^\theta \left(\theta=\pi\right)
    - P^0(\theta=0)\right]
   ={\beta\over 2} (b-a)={\beta r\over 2}\bigg|_a^b 
   \end{eqnarray}
   Adding the two contributions together, we get the correct result to be
    \begin{equation}
    {1\over 16\pi}\int \partial_c P^c  d^4x =  {1\over 4} \beta \left[- r^2 f' + 2r (1-2f)\right]_a^b
    \label{indvc}
    \end{equation}
    When evaluated on the Schwarzschild horizon  (with $\beta=8\pi M$ being the
    inverse temperature related to the periodicity in Euclidean time) the extra term $(\beta r/2)$ gives a contribution
    $\beta M=8\pi M^2$ which, when added to the original contribution $(-4\pi M^2)$ from the $P^r$,
    leads to a net contribution of $(+4\pi M^2)$. Thus the difference between the correct and wrong
    results in the case of Schwarzschild metric is just a flip of sign. It is easy to miss this or ``reinterpret"
    it, given the fact that one usually works in Euclidean sector.
    
    Unfortunately, this misses the correct interpretation. For Schwarzschild spacetime, the entropy
    and temperature are given by $S=4\pi M^2, T=\beta^{-1}=8\pi M$; so the combination
    $\beta F\equiv S-\beta E$ where $F$ is the free energy, is given by
    \begin{equation}
    S-\beta E=4\pi M^2 -(8\pi M) M=-4\pi M^2=-S
    \end{equation}
    The sign flip, arising due to the extra term makes all the difference between entropy $S$ and 
the free energy $S-\beta E$.

    To see this more clearly, let us consider a situation in which $f(r)$ has a simple zero at $r=r_H$
    with a finite derivative $|f'(r_H)|\equiv B$, say. This will lead to a compact horizon at
    $r=r_H$. Periodicity in the Euclidean time will now require
    $\beta B=4\pi$. The {\it wrong} expressions (\ref{wrongone},\ref{wrongtwo}) will give, on the horizon
    (where $f=0$)
    \begin{eqnarray}
    I_{\rm wrong}&=&-{1\over 4} r_H^2\beta f'(r_H)=-{1\over 4} r_H^2\beta B
    =-\pi r_H^2=-{1\over 4}({\rm Horizon Area})\nonumber \\
    &=&-S
    \end{eqnarray}
    But the correct result (\ref{indvc}) gives
    \begin{eqnarray}
    I_{\rm correct}&=&-{1\over 4} r_H^2\beta f'(r_H)+\beta(r_H/2)
    =-{1\over 4} r_H^2\beta B+\beta(r_H/2)\nonumber\\
    &=&-\pi r_H^2+\beta(r_H/2)
    =-(S-\beta E)
    \end{eqnarray} 
    with the energy associated with any horizon being given by $E=(r_H/2)$. 
    It is this interpretation and the possibility of defining the energy for any horizon
    which is missed when the wrong result is used.
    
    But how can a rigorously proved equation (\ref{pctok}) go wrong ? Actually, it did not. In the 
    summation on the right hand side of (\ref{pctok}), one also need to sum over a strange surface
    $\theta=$constant to get the correct result. To see this, note that the correct unit normal to the 
   $\theta=$constant surface is 
     \begin{eqnarray}
    n_\theta &=& {1\over \sqrt{|g^{\theta\theta}|}} (0,0,1,0) = r(0,0,1,0); \nonumber \\
     n^\theta &=& g^{\theta\theta} n_\theta = -{1\over r} (0,0,1,0)
    \end{eqnarray}
   The trace of the extrinsic curvature of this surface is
   \begin{eqnarray}
  K&=&  \nabla_a n^a = {1\over \sqrt{-g}} \partial_a ( \sqrt{-g} n^a) 
  = -{1\over r^2 \sin \theta} \partial_\theta \left( r^2 
    \sin\theta {1\over r}\right)\nonumber \\
    &=& -{1\over r} \cot\theta
   \end{eqnarray}
   On a   $\theta=$constant surface, the integral on the right hand side of (\ref{final}) will give
   \begin{eqnarray}
    {1\over 8\pi} \int_{\theta={\rm const}} d^3x \sqrt{h} K 
    &=&-{\beta\over 8\pi} \int_0^{2\pi} d\phi \int_a^b dr \left[\sqrt{f}{1\over \sqrt{f}}r\sin\theta\right]
    \left[{\cot\theta \over r}\right]\nonumber\\
    &=& -{\beta\over 4}(b-a)\cos\theta
    \end{eqnarray}
    When we evaluate this contribution at the surfaces $\theta=\pi$ and $\theta=0$ and
    subtract one from the other, we get
    \begin{equation}
    {1\over 8\pi} \int d^3x \sqrt{h} K |_{\theta=0}^{\theta=\pi}={\beta\over 2}(b-a)
    \end{equation}
    which is precisely the piece which was originally missing in action.
    
   The reason for this ``trouble" is that there are situations with integrable singularities for which 
   Gauss theorem is not applicable (or will give the wrong result) even in the ordinary 3-dimensional
   flat space, in standard spherical polar coordinates. Consider a vector field with components
   \begin{equation}
   v^r = A(r)  \sin\theta , \qquad  v^\theta = - \cos\theta , \qquad  v^\phi = 0
   \end{equation}
   with $A(r=0)=0$.  We want to integrate the quantity $\partial_\alpha v^\alpha$ over the 3 dimensional
   ball (${\cal B}$) of radius $R$. Since
   \begin{equation}
   \partial_\alpha v^\alpha = \left( A'(r) + 1 \right) \sin\theta 
   \end{equation}
   the integral is
   \begin{eqnarray}
   \int_0^Rdr \int_0^{2\pi} d\phi \int_0^{\pi} d\theta\, \partial_\alpha v^\alpha &=& 4\pi \int_0^R dr \, (A' + 1)
   \nonumber \\
   &=& 4\pi [A(R) + R ]
    \label{rteqn}
    \end{eqnarray}
   This is the correct result, obtained by inelegant index dynamics. We will now do it in a different way. Let us define another vector field $u^\alpha$ through the relation
   \begin{equation}
   d^3 x \, \left( \partial_\alpha v^\alpha\right)  = (\sqrt{g} d^3 x) \left[  {1\over \sqrt{g}} \partial_\alpha
    \left( \sqrt{g} u^\alpha\right) \right] = (\sqrt{g} d^3 x)\nabla \cdot {\bf u}
   \end{equation}
   The components of the new vector are given by
   \begin{equation}
   u^\alpha = {v^\alpha\over \sqrt{g}} = \left( {A(r) \over r^2}, - {1\over r^2} \cot \theta, 0 \right)
    \label{compo}
    \end{equation}
   The integral of $\partial_\alpha v^\alpha$ over $d^3x\equiv dr d\theta d\phi$
   is the same as the integral of the covariant divergence $\nabla \cdot {\bf u}$ over the 
   volume element $\sqrt{g} d^3x$. If we now use Gauss theorem, we get
    \begin{eqnarray}
    \int d^3x\, \partial_\alpha v^\alpha &=& \int_{\cal B} d^3x\, \sqrt{g}\,  \nabla\cdot {\bf u} = \int_{\partial {\cal B}}
    d^2x\, \sqrt{h}\,  {\bf n \cdot u}=4\pi R^2 u^r(R) \nonumber \\
     &=&  4\pi A(R)
     \label{wrongresult}
     \end{eqnarray}
    Only the radial component contributes and we get the {\it wrong result}.
    The second term $4\pi R$ of (\ref{rteqn}) is missing; the original analysis clearly shows
    that it originated from $(\partial v^\theta/\partial \theta)$ which we miss
    unless we deal with $\theta =$ constant ``surfaces''. How can an old faithful
    like Gauss theorem lead us astray ?
    Before applying Gauss theorem to the vector field ${\bf u}$ in (\ref{wrongresult})
    we need to ensure that the vector field is not singular. Equation (\ref{compo}) clearly
    shows that the $u^\theta$ component diverges on the $z-$axis corresponding to
    $\theta =(0,\pi)$. There is a  singularity on this axis which gives
    the extra contribution. 
   
In summary, we get the correct interpretation of the action as the free energy of horizon, when the surface terms are taken into account correctly.

    \end{document}